# INVESTIGATION OF THE ROLE OF THE ROENTGEN INTERACTION IN ANGULAR MOMENTUM TRANSFER WITHIN AN ELECTRIC BOSE EINSTEIN CONDENSATE

L G Bousiakou[1,2], O M Aldossary[1], W S Almarhoon[1]

[1]Department of Physics and Astronomy, College of Science, King Saud University, Riyadh, Saudi Arabia

[2]Department of Automation Engineering, Technological Educational Institute of Piraeus, Egaleo Athens, 12244, Greece

**Abstact**

The exact action of the Roentgen effect at an atomic level is investigated within the context of vortex formation using two Laguerre Gaussian beams. Starting from a modified Gross-Pitaevskii equation in the dipole approximation that takes into account the coupling of the atomic system to the electromagnetic fields the Röntgen interaction term is introduced and viewed within the context of such a Raman transition. A rigorous investigation of the corresponding Rabi frequency reveals that the Roentgen term acts at different levels compared to the –d.E term and thus would not cause any changes in a fully quantum mechanical treatment.

## 1. Introduction

The interaction of light, carrying quantized orbital angular momentum[1] (OAM), with atomic or molecular matter is of considerable research interest [2-4]. This quest is increasingly becoming feasible due to the progress that has been made in creation [5-7], manipulation[8], detection[9,10], and application[11] of the OAM states of light.

Recent theoretical studies of the interaction of Laguerre Gaussian (LG) beams with atoms within Bose Einstein Condensates (BECs) have been conducted[12-14] examining how the main features of Doppler cooling and trapping are modified when a plane wave or a fundamental Gaussian beam is replaced with Laguerre Gaussian light. It appears reasonable that the interaction of atoms with beams possessing OAM should lead to new effects. It should exhibit novel rotational effects, in addition to the expected translational effects normally encountered. The effects involve changes both to the internal and gross motion of the atom.

The first generation of a vortex in a Bose Einstein Condensate used a "phase engineering" scheme involving a rapidly rotating Gaussian laser beam coupling the external motion to internal state Rabi oscillations[15,16]. Later schemes included mechanically stirring the BEC with a focused laser beam[17] and "phase imprinting" by adiabatic passage[18,19]. However, transfer of OAM from the rotating light beams in these earlier schemes is not well-defined[20]. Over the past decade, numerous papers[21-24] proposed generating vortices in a BEC using stimulated Raman processes with configurations of optical fields that provide OAM, such as LG beams.

In this work we consider vorticity effects induced by LG beams in an electric BEC. In particular we consider the case of two co-propagating $\sigma^+$ and $\sigma^-$ polarized laser beams along the *z* direction (gravity's direction) to induce a Raman transition between two hyperfine levels of ground-state alkali atoms.

An electric BEC is described by an appropriately modified Gross Pitaevskii equation ( GPE) in the presence of electromagnetic fields, where the constituent atoms are characterized by an electric dipole moment. Within this frame the inclusion of the Roentgen term ensures the proper treatment, allowing momentum changes to be well determined when atoms interact which any given state of a radiation field.

## 2. The Raman vortex coupler

It has been proposed by Marzlin et al [12] that a Bose Einstein condensate ( BEC) can be spun into a vortex state using a coherent coupler. Considering that such a condensate is formed by alkali atoms, the coupler which can induce a Raman transition between two hyperfine levels of the ground state of the ultracold atoms using two co-propagating $\sigma^+$ and $\sigma^-$ polarized Laguerre-Gaussian laser beams with different frequencies along the z direction.

In this case a modified non-linear Schroedinger equation in the dipole approximation can be written to describe the coupling between the condensate wave functions $\Psi_-$ and $\Psi_+$ corresponding to the magnetic sublevels $|-\rangle = |F=1, M_F=-1\rangle$ and $|+\rangle = |F=1, M_F=1\rangle$ respectively. It is in essence a Gross-Pitaevskii equation that is expressed as:

$$i\hbar\dot{\Psi}_- = (\frac{P^2}{2M} - E_\mu)\Psi_- + (\frac{d^2}{\varepsilon_0} + U_0)N\left[2|\Psi_-|^2 + |\Psi_+|^2\right]\Psi_- + \frac{\hbar|\Omega_1^{(-)}|^2}{4\Delta\omega}\Psi_-$$
$$+ \frac{\hbar\Omega_1^{(-)}\Omega_2^{(+)}}{4\Delta\omega}e^{-i\Delta\omega t}\Psi_+ \quad (2.1)$$
$$i\hbar\dot{\Psi}_+ = (\frac{P^2}{2M} - E_\mu)\Psi_+ + (\frac{d^2}{\varepsilon_0} + U_0)N\left[2|\Psi_+|^2 + |\Psi_-|^2\right]\Psi_+ + \frac{\hbar|\Omega_2^{(-)}|^2}{4\Delta\omega}\Psi_+$$
$$+ \frac{\hbar\Omega_1^{(+)}\Omega_2^{(-)}}{4\Delta\omega}e^{i\Delta\omega t}\Psi_-$$

We note that $P^2/2M$ is a kinetic energy term and $E_\mu$ the chemical potential. We identify $N$ to be the number of atoms in the ensemble and $d$ the electric dipole moment. The expression:

$$N\frac{d^2}{\varepsilon_0}|\psi_\pm|^2 \quad (2.2)$$

represents an effective dipole-dipole interaction term [25].

Additionally the two Raman beams are detuned by a frequency $\Delta\omega = \omega_2 - \omega_1$ and $\Omega_i^+$ represents the positive frequency part of the corresponding Rabi frequencies related to the two Raman beams. Furthermore the last two terms of equation (2.1) represent the Raman coupling which can coherently transfer the initial non-rotating condensate in level $|-\rangle$ to the initially empty vortex state level $|+\rangle$.

We note that gravity effects are excluded here since we are interested only in the rotational motion of the condensates in the *x-y* plane.

In this case it has been shown by Marzlin et al [12] that a population transfer of 50% to the VS is possible for fixed values of $\Delta\omega$, which is possible to circumvent by the introduction of a linearly time-dependent frequency difference between the two lasers.

Nevertheless within this frame the role of the Roentgen term in vortex nucleation has not been explored and a rigorous treatment is employed in order to see it's effect within such a system.

### 3. The Röntgen term and the modified Rabi frequency in the case of an electric BEC

We define the Rabi frequency $\Omega_1^{(+)}$ for the $\sigma^+$ polarized beam between a final $|\Psi_f\rangle$ and initial level $|\Psi_i\rangle$, such as:

$$\Omega_1^+ = \frac{\langle\Psi_f|V_{em}|\Psi_i\rangle}{\hbar} \quad (3.1)$$

According to Bousiakou et al [25] the term $V_{em}$ is a modified electromagnetic interaction term that provides a full description of the electric and magnetic dipole interactions within our system such that:

$$V_{em} = -\vec{d}\cdot\vec{E}(\vec{r}) - \vec{\mu}\cdot\vec{B}(\vec{r}) + \frac{1}{2M}\left[\vec{P}\cdot\left\{\vec{d}\times\vec{B}(\vec{r}) - \frac{1}{c^2}\vec{\mu}\times\vec{E}(\vec{r})\right\} + \left\{\vec{d}\times\vec{B}(\vec{r}) - \frac{1}{c^2}\vec{\mu}\times\vec{E}(\vec{r})\right\}\cdot\vec{P}\right]$$

(3.2)

Where $\mu$ is the magnetic dipole moment.

The terms $\vec{P}\cdot(\vec{d}\times\vec{B}(\vec{r}))$ and $\vec{P}\cdot(\vec{\mu}\times\vec{E}(\vec{r}))$ are identified as the Roentgen interaction[25, 26-29] and the Aharonov Casher terms[25,29,30] respectively.

The *electric BEC* is defined as the case where the constituent atoms are characterized by an electric dipole moment only, thus the above term reduces to:

$$V_{em} = -\vec{d}\cdot\vec{E}(\vec{r}) + \frac{1}{2M}\left[\vec{P}\cdot\left\{\vec{d}\times\vec{B}(\vec{r})\right\} + \left\{\vec{d}\times\vec{B}(\vec{r})\right\}\cdot\vec{P}\right]$$

(3.3)

, which can be re-written as

$$V_{em} = -\vec{d}\cdot\vec{E}(\vec{r}) + \left[\vec{d}\cdot\left\{\vec{B}(\vec{r})\times\vec{v}(\vec{r})\right\}\right]$$

(3.4)

after some manipulations. The term $\vec{v}(r)$ represents the BEC velocity profile.

In order to calculate the Rabi frequency related to the Röntgen effect, we employ Fermi's golden rule where the internal coordinates are spherical and the wave functions in the matrix elements in the basis of hydrogenic wave functions, considering our initial assumption that the BECs under consideration is primarily formed using alkali, hydrogen like atoms, such as $Rb^{87}$, $Na^{23}$, etc.

The normalized hydrogenic wave functions[31] are thus given by :

$$\Psi_{nlm} = \sqrt{\left(\frac{2}{na}\right)^3 \frac{(n-l-1)!}{2n[(n+l)!]^3}} e^{-r/na}\left(\frac{2r}{na}\right)^l \left[L_{n-l-1}^{2l+1}\right] Y_l^m(\theta,\phi)$$

(3.5)

where $L_{n-l-1}^{2l+1}$ is an associated Laguerre polynomial and $Y_l^m$ is the spherical harmonics, defined by:

$$Y_l^m(\theta,\phi) = \varepsilon\sqrt{\frac{(2l+1)}{4\pi}\frac{(l-|m|)!}{(l+|m|)!}} e^{im\phi} P_l^m(\cos\theta)$$

(3.6)

The first few spherical harmonics and associate Laguerre polynomials, are listed in Tables 1 and 2 below, i.e. :

| $Y_0^0 = \left(\frac{1}{4\pi}\right)^{1/2}$ | $Y_2^{\pm 2} = \left(\frac{15}{32\pi}\right)^{1/2}\sin^2\theta e^{\pm 2i\phi}$ |
|---|---|
| $Y_1^0 = \left(\dfrac{3}{4\pi}\right)^{1/2} \cos\theta$ | $Y_3^0 = \left(\dfrac{7}{16\pi}\right)^{1/2} \left(5\cos^3\theta - 3\cos\theta\right)$ |
| $Y_1^{\pm 1} = \mp\left(\dfrac{3}{8\pi}\right)^{1/2} \sin\theta e^{\pm i\phi}$ | $Y_3^{\pm 1} = \mp\left(\dfrac{21}{64\pi}\right)^{1/2} \sin\theta\left(5\cos^2\theta - 1\right)e^{\pm i\phi}$ |
| $Y_2^0 = \left(\dfrac{5}{16\pi}\right)^{1/2} \left(3\cos^2\theta - 1\right)$ | $Y_3^{\pm 2} = \left(\dfrac{105}{32\pi}\right)^{1/2} \sin^2\theta\cos\theta e^{\pm 2i\phi}$ |
| $Y_2^{\pm 1} = \mp\left(\dfrac{5}{16\pi}\right)^{1/2} \sin\theta\cos\theta e^{\pm i\phi}$ | $Y_3^{\pm 3} = \mp\left(\dfrac{35}{64\pi}\right)^{1/2} \sin^3\theta e^{\pm 3i\phi}$ |

**Table 1.** The first few spherical harmonics[31], $Y_l^m(\theta,\phi)$

| | |
|---|---|
| $L_0^0 = 1$ | $L_0^2 = 2$ |
| $L_1^0 = -x + 1$ | $L_1^2 = -6x + 18$ |
| $L_2^0 = x^2 - 4x + 2$ | $L_2^2 = 12x^2 - 96x + 144$ |
| $L_0^1 = 1$ | $L_0^3 = 6$ |
| $L_1^1 = -2x + 4$ | $L_1^3 = -24x + 96$ |
| $L_2^1 = 3x^2 - 18x + 18$ | $L_1^3 = 60x^2 - 600x + 1200$ |

**Table 2.** The first few associate Laguerre polynomials[31]

Using equations (3.1) and (3.4) we can write the Rabi frequency for the $\sigma^+$ polarized beam as:

$$\Omega_1^+ = \frac{\langle\Psi_f|-\vec{d}\cdot\vec{E}(\vec{r})|\Psi_i\rangle}{\hbar} + \frac{\langle\Psi_f|\vec{d}\cdot\{\vec{B}(\vec{r})\times\vec{v}(r)\}|\Psi_i\rangle}{\hbar} \quad (3.7)$$

We calculate the matrix elements of these two terms using the hydrogen wave functions for n=3 which are given in Table 3, below:

| n | l | m | $\psi_{nlm} = R_{nl}Y_{lm}$ |
|---|---|---|---|
| 3 | 0 | 0 | $\dfrac{1}{81\sqrt{3\pi}}\left(\dfrac{1}{a_o}\right)^{3/2}\left(27 - 18\dfrac{r}{a_o} + 2\left(\dfrac{r}{a_o}\right)^2\right)e^{-r/a_o}$ |
| 3 | 1 | 0 | $\dfrac{1}{81}\sqrt{\dfrac{2}{\pi}}\left(\dfrac{1}{a_o}\right)^{3/2}\left(6 - \dfrac{r}{a_o}\right)\dfrac{r}{a_o}e^{-r/a}\cos\theta$ |

| 3 | 1 | ±1 | $\frac{1}{8\sqrt{\pi}}\left(\frac{1}{a_o}\right)^{3/2}\left(6-\frac{r}{a_o}\right)\frac{r}{a_o}e^{-r/3a}\sin\theta\,e^{\pm i\phi}$ |
| --- | --- | --- | --- |
| 3 | 2 | 0 | $\frac{1}{81\sqrt{6\pi}}\left(\frac{r}{a_o}\right)^2 e^{-r/3a}(3\cos^2\theta-1)$ |
| 3 | 2 | ±1 | $\frac{1}{81\sqrt{6\pi}}\left(\frac{r}{a_o}\right)^2 e^{-r/3a}\sin\theta\cos\theta\,e^{\pm i\phi}$ |
| 3 | 2 | ±2 | $\frac{1}{162\sqrt{\pi}}\left(\frac{r}{a_o}\right)^2 e^{-r/3a}\sin^2\theta\,e^{\pm 2i\phi}$ |

**Table 3:** The first few wave functions[31] of hydrogen atom in the third shell, n=3.

In general the electric field of a linearly polarized Laguerre-Gaussian beam propagating in the z-direction is given in cylindrical coordinates by:

$$\vec{E}(\vec{r}) = E_0 e^{-r^2/w^2}\left(\frac{\sqrt{2}r}{w}\right)^l e^{il\phi}e^{ikz}\vec{x} \qquad (3.8)$$

while the magnetic field in the linear polarization given as (SI units)

$$\vec{B}(\vec{r}) = \hat{z}\times\vec{E}(\vec{r})/c \qquad (3.9)$$

where

$$\vec{B}(\vec{r}) = \hat{z}\times(E_0/c)e^{-r^2/w^2}\left(\frac{\sqrt{2}r}{w}\right)^l e^{il\phi}e^{ikz}\hat{x} \qquad (3.10)$$

or

$$\vec{B}(\vec{r}) = B_0 e^{-r^2/w^2}\left(\frac{\sqrt{2}r}{w}\right)^l e^{il\phi}e^{ikz}\vec{y} \qquad (3.11)$$

where $l$ is the azimuthal index of the beam and $r^2 = x^2 + y^2$. Similar expressions applies to the $\sigma^-$ polarized beam with $\phi$ replaced by $-\phi$ in the azimuthal angular dependence term of the beam, i.e. $\exp(-il\phi)$.

Within this context we express equation (4.3) such as:

$$\Omega_1^+ = \frac{-\langle\Psi_f|eE_0 r\sin\theta\cos\phi|\Psi_i\rangle}{\hbar} + \frac{\langle\Psi_f|eB_0 v_0 r\cos\theta\cos\phi|\Psi_i\rangle}{\hbar}$$

The corresponding matrix elements are calculated to be:

$$-eE_0\langle 310|r\sin\theta\cos\phi|300\rangle e^{-r^2/w^2}\left(\frac{\sqrt{2}r}{w}\right)^l e^{il\phi}e^{ikz}=0 \quad (3.12)$$

$$-eE_0\langle 31\pm 1|r\sin\theta\cos\phi|300\rangle e^{-r^2/w^2}\left(\frac{\sqrt{2}r}{w}\right)^l e^{il\phi}e^{ikz}\approx(60.4052856 a_0 eE_0)e^{-r^2/w^2}\left(\frac{\sqrt{2}r}{w}\right)^l e^{il\phi}e^{ikz}$$

$$(3.13)$$

where $a_0 = 5.2917720859(36) \times 10^{-11}$ m is the Bohr radius of hydrogen and $e = 1.6 \times 10^{-19}$ Cb is the electron charge.

Additionally:

$$-eE_0\langle 320|r\sin\theta\cos\phi|300\rangle e^{-r^2/w^2}\left(\frac{\sqrt{2}r}{w}\right)^l e^{il\phi}e^{ikz}=0 \quad (3.14)$$

$$-eE_0\langle 32\pm 1|r\sin\theta\cos\phi|300\rangle e^{-r^2/w^2}\left(\frac{\sqrt{2}r}{w}\right)^l e^{il\phi}e^{ikz}=0 \quad (3.15)$$

As shown above all of them are zero except $\langle 31\pm 1|-\vec{d}\cdot\vec{E}(\vec{r})|300\rangle$, where $\Delta l = 1$ and $\Delta m = \pm 1$

The matrix elements of Röntgen $\vec{d}\cdot(\vec{B}(\vec{r})\times\vec{v}(\vec{r}))$ term are calculated to be:

$$eB_0 v_0\langle 310|r\cos\theta\cos\phi|300\rangle e^{-r^2/w^2}\left(\frac{\sqrt{2}r}{w}\right)^l e^{il\phi}e^{ikz}=0 \quad (3.16)$$

$$eB_0 v_0\langle 31\pm 1|r\cos\theta\cos\phi|300\rangle e^{-r^2/w^2}\left(\frac{\sqrt{2}r}{w}\right)^l e^{il\phi}e^{ikz}=0 \quad (3.17)$$

$$eB_0 v_0\langle 320|r\cos\theta\cos\phi|300\rangle e^{-r^2/w^2}\left(\frac{\sqrt{2}r}{w}\right)^l e^{il\phi}e^{ikz}=0 \quad (3.18)$$

$$eB_0 v_0\langle 32\pm 1|r\cos\theta\cos\phi|300\rangle e^{-r^2/w^2}\left(\frac{\sqrt{2}r}{w}\right)^l e^{il\phi}e^{ikz}\approx(0.8118988\pi ea_0 B_0 v_0)e^{-r^2/w^2}\left(\frac{\sqrt{2}r}{w}\right)^l e^{il\phi}e^{ikz}$$

$$(3.19)$$

As shown above all of them are zero except when $\langle 32\pm 1|\vec{d}\cdot(\vec{B}(r)\times\vec{v}(\vec{r}))|300\rangle$ where $\Delta l = 2$ and $\Delta m = \pm 1$ considering the 2-photon nature of the Raman process.

Thus following the above detailed formalism, we conclude from equations (3.13) and (3.19) that the Röntgen term and the $-\vec{d}\cdot\vec{E}(\vec{r})$ term act between different levels, such as:

$$\Omega^+_{1\ -\vec{d}\cdot\vec{E}} = \frac{\langle\Psi_f(3p)|-\vec{d}\cdot\vec{E}(\vec{r})|\Psi_i(3s)\rangle}{\hbar} \qquad (3.20)$$

and

$$\Omega^+_{1\ Rontgen\_term} = \frac{\langle\Psi_f(3d)|\vec{d}\cdot\{\vec{B}(\vec{r})\times\vec{v}(\vec{r})\}|\Psi_i(3s)\rangle}{\hbar} \qquad (3.21)$$

Thus the Roentgen term would have no direct effect on the vorticity in such a Raman coupler. This gives us a clear insight on the role of the Roentgen term for the quantum optics perspective and the details of its interaction within a BEC under influence of Laguerre Gaussian beams.

**4. Conclusions**

The exact action of the Roentgen effect at an atomic level is investigated within the context of vortex formation using two Laguerre Gaussian (LG) beams. Our treatment reveals that due to the fact that the Roentgen term acts at different levels compared to the –d.E term it would not cause any changes in a fully quantum mechanical treatment.

**Acknowledgements**

We are grateful to Dr. C. R. H. Bennett for helpful discussions and clarification of issues discussed in this paper. This work was supported by NSTIP strategic technologies programs, number (11-MAT-1898-02) in the Kingdom of Saudi Arabia.